%% file: RK.tex
\documentclass[review,onefignum,onetabnum]{siamart171218}

\input{ex_shared}
\input{tcilatex}

\begin{document}
	
	\maketitle
	
	\begin{abstract}
		It is demonstrated in this paper that the propagation of the electric wave
		field in a heterogeneous medium in 3D can sometimes be governed well by a
		single PDE, which is derived from the Maxwell's equations. The corresponding
		component of the electric field dominates two other components. This
		justifies some past results of the second author with coauthors about
		numerical solutions of coefficient inverse problems with experimental
		electromagnetic data. In addition, since it is simpler to work in
		applications with a single PDE rather than with the complete Maxwell's
		system, then the result of this paper might be useful to researchers working
		on applied issues of the propagation of electromagnetic waves in
		inhomogeneous media.
	\end{abstract}
	
	
	
	\section{Introduction}
	
	\label{sec:1}
	
	In some previous works of the second author with coauthors coefficient
	inverse problems were solved for frequency dependent microwave experimental
	electromagnetic data using only the single Helmholtz equation, see, e.g. 
	\cite{Khoa2,Khoa3,KlibKol4}. Reconstruction results were quite accurate
	ones. A similar observation took place in \cite{BK,Thanh}, although for time
	dependent experimental data. Thus, a natural question to pose is:\emph{\
		Given that the propagation of the electromagnetic wave field is governed by
		the Maxwell's equations, why the use of only a single equation has provided
		accurate reconstruction results?} A positive heuristic answer to this
	question can be found in the classical textbook of M. Born and E. Wolf \cite[%
	pages 695,696]{BW} for the frequency domain case. In addition, this question
	was positively addressed numerically in \cite{BMM} for the time domain case
	and in \cite{KNN} for the frequency domain case. It was demonstrated
	computationally in \cite{BMM,KNN} that if the incident electric wave field
	has only a single non zero component, then this component dominates two
	other components while propagating through the medium, and its propagation
	is well governed by a wave-like PDE. That PDE is either the Helmholtz
	equation in the frequency domain or the corresponding hyperbolic equation in
	the time domain.
	
	The goal of this paper is to investigate the above question rigorously. We
	believe that the results of this paper might be useful not only for an
	analytical explanation of the accuracy of imaging results of \cite%
	{Khoa2,Khoa3,KlibKol4} but also for applied mathematicians, physicists and
	engineers working on various topics of electromagnetic waves propagation.
	Indeed, it is clear that it is easier to work in applications with a single
	PDE rather than with the whole Maxwell's system.
	
	In section 2 we work in time domain. These results are used then in section
	3, where we derive our desired conclusion for the case of the frequency
	domain. In section 4, we link our main Theorem 2 of section 3 with the above
	cited results of \cite{Khoa2,Khoa3,KlibKol4}. In section 5 (Appendix) we
	prove a certain energy estimate.
	
	\section{Time Domain}
	
	\label{sec:2}
	
	Consider the Maxwell's equations in a non magnetic medium 
	\begin{equation}
		\mathrm{curl}\,\mathbf{H}=\varepsilon (\mathbf{x})\mathbf{E}_{t},~\mathrm{%
			curl}\,\mathbf{E}=-\mathbf{H}_{t},~\mathrm{div}\,\mathbf{H}=0,~\mathbf{x}\in 
		\mathbb{R}^{3},t>0,  \label{2.1}
	\end{equation}%
	where $\varepsilon (\mathbf{x})$ is the spatially distributed dielectric
	constant. We work in this paper with dimensionless variables, since
	variables were made dimensionless in the above cited works about the
	experimental data. Thus, in vacuum $\varepsilon (\mathbf{x})=1,$ and we also
	assume that the magnetic permeability $\mu \equiv 1$. Let $\mathbf{\nu }\in 
	\mathbb{S}^{2}=\left\{ \left\vert \mathbf{\nu }\right\vert =1\right\} $ be a
	unit vector of the direction of propagation of the incident electric wave
	field. If the space $\mathbb{R}^{3}$ is vacuum, then equations \vspace{-3mm}(%
	\ref{2.1}) admit the following solution \cite{R}: 
	\begin{equation}
		\mathbf{E}^{0}(\mathbf{x},t)=\mathbf{j}\delta (t+t_{0}-\mathbf{x}\cdot 
		\mathbf{\nu }),~\mathbf{H}^{0}(\mathbf{x},t)=\left( \mathbf{\nu }\times 
		\mathbf{j}\right) \delta (t+t_{0}-\mathbf{x}\cdot \mathbf{\nu }),
		\label{2.2}
	\end{equation}%
	where $\delta (t)$ is the Dirac delta function, $t_{0}$ is an arbitrary
	number, $\,\mathbf{x}\cdot \mathbf{\nu }$ denotes the scalar product of
	these two vectors and $\mathbf{j}\cdot \mathbf{\nu }=0$. The vector $\mathbf{%
		j}$ defines the polarization of this wave, 
	\begin{equation}
		\mathbf{j}=\left( j_{1},j_{2},j_{3}\right) ,  \label{1}
	\end{equation}
	where $j_{1},j_{2},j_{3}$ are some constants. In the sequel we assume that $%
	\mathbf{j}\in \mathbb{S}^{2}$. The orthogonality of vectors $\mathbf{j}$ and 
	$\mathbf{\nu }$ is necessary to satisfy the equation $\mathrm{div}\mathbf{H}%
	=0$.
	
	Below $\mathbf{j}$ and $\mathbf{\nu }$ are assumed to be arbitrary but fixed
	vectors. Therefore we do not indicate dependence of the solution and some
	functions on these parameters for brevity, unless this is really necessary.
	Still, we use the parameter $\mathbf{\nu }$ to indicate some domains and a
	plane wave for a more clear understanding. Below vectors $\mathbf{j}$, $%
	\mathbf{\nu }$, $\mathbf{E}$, $\mathbf{H}$, etc. are row vectors, see, e.g. (%
	\ref{1}).
	
	Let $R>0$ be an arbitrary number. Consider the ball $B$ with the center at $%
	\left\{ 0\right\} $ and the radius $R$. Let the sphere $S=\partial B.$ Then%
	\begin{equation*}
		B=\{\mathbf{x}\in \mathbb{R}^{3}:\,|\mathbf{x}|<R\},~S=\{\mathbf{x}\in 
		\mathbb{R}^{3}:\,|\mathbf{x}|=R\}.
	\end{equation*}%
	We assume that 
	\begin{equation}
		1\le\varepsilon (\mathbf{x})\le \varepsilon_0 \>\text{ in }B\text{, }%
		\varepsilon (\mathbf{x})=1\text{ in }\mathbb{R}^{3}\diagdown B,\text{ }
		\label{2.20}
	\end{equation}%
	where $\varepsilon_0\ge1$ is a constant, and 
	\begin{equation}
		t_{0}=\min_{\mathbf{x}\in S}(\mathbf{x}\cdot \mathbf{\nu })=-R.  \label{2.21}
	\end{equation}
	
	Let the incident plane wave propagates in the vacuum for $t<0$ and meets the
	heterogeneous ball $B$ at a moment of time $t=0$. Then the propagation of
	the electromagnetic wave field is governed by the following Cauchy problem:%
	\begin{equation}
		\func{curl}\mathbf{H}=\varepsilon \left( \mathbf{x}\right) \mathbf{E}_{t},~%
		\func{curl}\mathbf{E}=-\mathbf{H}_{t},~\left( \mathbf{x},t\right) \in 
		\mathbb{R}^{4},  \label{2.3}
	\end{equation}%
	\begin{equation}
		\mathbf{E}|_{t<0}=\mathbf{E^{0}}(\mathbf{x},t),~\mathbf{H}|_{t<0}=\mathbf{%
			H^{0}}(\mathbf{x},t).  \label{2.4}
	\end{equation}%
	For the sake of convenience, we reduce now problem (\ref{2.3}), (\ref{2.4})
	to the case when only the vector function $\mathbf{E}(\mathbf{x},t)$ is
	unknown. Since $\func{div}\left( \func{curl}\mathbf{U}\right) =0$ for any
	appropriate vector function $\mathbf{U,}$ then applying the operator $\func{%
		div}$ to both sides of the first equation (\ref{2.3}), we obtain 
	\begin{equation}
		\func{div}\left( \varepsilon \left( \mathbf{x}\right) \mathbf{E}_{t}\right)
		=0.  \label{2.5}
	\end{equation}%
	Integrating \vspace{-3mm}(\ref{2.5}) with respect to $t$ and using (\ref{2.4}%
	), we obtain 
	\begin{equation}
		\func{div}\left( \varepsilon \left( \mathbf{x}\right) \mathbf{E}(\mathbf{x}%
		,t)\right) -\func{div}\left( \varepsilon \left( \mathbf{x}\right) \mathbf{%
			E^{0}}\left( \mathbf{x},\mathbf{\nu ,}0\right) \right) =0.  \label{2.6}
	\end{equation}%
	Note that $\mathrm{div}\,\mathbf{E}^{0}(\mathbf{x},t)=0$, since $\mathbf{j}%
	\cdot \mathbf{\nu }=0$. Indeed, 
	\begin{equation*}
		\mathrm{div}\,\mathbf{E}^{0}(\mathbf{x},t)=-(\mathbf{j}\cdot \mathbf{\nu }%
		)\delta ^{\prime }(t+t_{0}-\mathbf{x}\cdot \mathbf{\nu })=0.
	\end{equation*}%
	Hence, 
	\begin{equation}
		\mathrm{div}\,(\varepsilon (\mathbf{x})\mathbf{E}^{0}(\mathbf{x},0))=\nabla
		\varepsilon (\mathbf{x})\cdot \mathbf{E}^{0}(\mathbf{x},0)=(\nabla
		\varepsilon (\mathbf{x})\cdot \mathbf{j})\delta (t_{0}-\mathbf{x}\cdot 
		\mathbf{\nu })=0.  \label{2.7}
	\end{equation}%
	This is because by (\ref{2.21}) $\text{supp}\>\{\delta (t_{0}-\mathbf{x}%
	\cdot \mathbf{\nu })\}$ is the tangent plane to $S$, namely $(\mathbf{x}%
	\cdot \mathbf{\nu })=-R$, along which $\nabla \varepsilon (\mathbf{x})=0$.
	Hence $\mathrm{div}\,(\varepsilon (\mathbf{x})\mathbf{E}(\mathbf{x},t))=0$
	for any $t$. Hence, problem (\ref{2.3}), (\ref{2.4}) is reduced to the
	following problem with $n(\mathbf{x})=\sqrt{\varepsilon (\mathbf{x})}:$ 
	\begin{equation}
		n^{2}(\mathbf{x})\mathbf{E}_{tt}-\Delta \mathbf{E}-\nabla (\mathbf{E}\cdot
		\nabla \ln n^{2}(\mathbf{x}))=0,~(\mathbf{x},t)\in \mathbb{R}^{4},~
		\label{2.8}
	\end{equation}%
	\begin{equation}
		\mathbf{E}|_{t<0}=\mathbf{E}^{0}(\mathbf{x},t).  \label{2.9}
	\end{equation}
	
	Define two domains $D_{-}\left( \mathbf{\nu }\right) $ and $D_{+}\left( 
	\mathbf{\nu }\right) $ as%
	\begin{equation*}
		D_{-}\left( \mathbf{\nu }\right) =\{\mathbf{x}\in \mathbb{R}^{3}:\,\mathbf{x}%
		\cdot \mathbf{\nu }+R<0\},
	\end{equation*}%
	\begin{equation*}
		D_{+}\left( \mathbf{\nu }\right) =\{\mathbf{x}\in \mathbb{R}^{3}:\mathbf{x}%
		\cdot \mathbf{\nu }+R\geq 0\}.
	\end{equation*}%
	Note that the domain $D_{-}\left( \mathbf{\nu }\right) $ is situated outside
	of $B$, while $B\subset D_{+}\left( \mathbf{\nu }\right) $.
	
	To define geodesic lines, we partially follow our paper \cite{KR1}. The
	function $n(\mathbf{x})$ generates the Riemannian metric 
	\begin{equation*}
		d\tau =n(\mathbf{x})\left\vert d\mathbf{x}\right\vert ,\text{ }|d\mathbf{x}|=%
		\sqrt{(dx_{1})^{2}+(dx_{2})^{2}+(dx_{3})^{2}}.
	\end{equation*}%
	For each vector $\mathbf{\nu }\in \mathbb{S}^{2}$ define the plane $\Sigma (%
	\mathbf{\nu })$ as%
	\begin{equation}
		\Sigma (\mathbf{\nu })=\{\mathbf{\xi }\in \mathbb{R}^{3}:\mathbf{\>\xi \cdot
			\nu }=-R\}.  \label{3.1}
	\end{equation}%
	Observe that the plane $\Sigma (\mathbf{\nu })$ is tangent to $S$ at the
	point $\mathbf{\xi }_{\tan }=-R\mathbf{\nu }$. Hence, $\Sigma (\mathbf{\nu }%
	)\cap B=\varnothing $ and the vector $\mathbf{\nu }$ is a normal vector to
	the plane $\Sigma (\mathbf{\nu }).$ Consider an arbitrary point $\mathbf{y}%
	\in \Sigma (\mathbf{\nu }).$ This point can be represented as 
	\begin{equation}
		\mathbf{y}=\mathbf{y}(a_{2},a_{3})=-R\mathbf{\nu }+a_{2}\mathbf{e}_{2}+a_{3}%
		\mathbf{e}_{3},\quad (a_{2},a_{3})\in \mathbb{R}^{2},  \label{3.2}
	\end{equation}%
	where unit vectors $\mathbf{\nu }$, $\mathbf{e}_{2}=\mathbf{j}$, $\mathbf{e}%
	_{3}=\mathbf{\nu }\times \mathbf{j}$ form an orthogonal triple. Note that
	vectors $\mathbf{e}_{2}$, $\mathbf{e}_{3}$ are parallel to the plane $\Sigma
	(\mathbf{\nu }).$
	
	Let the function $\varphi (\mathbf{x},\mathbf{\nu })$ be the solution of the
	Cauchy problem for the eikonal equation, 
	\begin{equation}
		|\nabla _{\mathbf{x}}\varphi (\mathbf{x},\mathbf{\nu })|^{2}=n^{2}(\mathbf{x}%
		),\quad \varphi (\mathbf{x},\mathbf{\nu })|_{\mathbf{x}\in \Sigma (\mathbf{%
				\nu })}=0  \label{3.3}
	\end{equation}%
	satisfying the following conditions: 
	\begin{equation}
		\varphi (\mathbf{x},\mathbf{\nu })\left\{ 
		\begin{array}{c}
			<0\text{ if }\mathbf{x}\in D_{-}\left( \mathbf{\nu }\right) , \\ 
			>0\text{ if }\mathbf{x}\in D_{+}\left( \mathbf{\nu }\right) .%
		\end{array}%
		\right.  \label{3.30}
	\end{equation}%
	The number $|\varphi (\mathbf{x},\mathbf{\nu })|$ is the Riemannian distance
	between the point $\mathbf{x}$ and the plane $\Sigma (\mathbf{\nu })$. From
	the Physics standpoint, $|\varphi (\mathbf{x},\mathbf{\nu })|$ is the travel
	time between the point $\mathbf{x}$ and the plane $\Sigma (\mathbf{\nu })$.
	For $\mathbf{\xi }\cdot \mathbf{\nu }<-R,$ i.e. in the domain $D_{-}\left( 
	\mathbf{\nu }\right) $, the function $\varphi (\mathbf{x},\mathbf{\nu })$
	has the form $\varphi (\mathbf{x},\mathbf{\nu })=\mathbf{x}\cdot \mathbf{\nu 
	}+R$. To find the function $\varphi (\mathbf{x},\mathbf{\nu })$ in the
	domain $D_{+}\left( \mathbf{\nu }\right) ,$ we need to solve problem (\ref%
	{3.3}), (\ref{3.30}) in this domain. It is known that to do this, we need to
	solve the following Cauchy problem for a system of ordinary differential
	equations \cite{R2}: 
	\begin{equation}
		\frac{d\mathbf{x}}{ds}=\frac{p(\mathbf{x,\nu })}{n^{2}(\mathbf{x})},\quad 
		\frac{d\mathbf{p}(\mathbf{x,\nu })}{ds}=\nabla \ln n(\mathbf{x}), \quad 
		\frac{d\varphi (\mathbf{x,\nu })}{d s}=1,s>0,  \label{3.4}
	\end{equation}%
	\begin{equation}
		\mathbf{x}|_{s=0}=\mathbf{y},~\mathbf{p}|_{s=0}=\mathbf{\nu },~\varphi
		|_{s=0}=0,  \label{3.5}
	\end{equation}%
	where $\mathbf{y}\in \Sigma (\mathbf{\nu })$ is an arbitrary point of the
	plane $\Sigma (\mathbf{\nu }),$ see (\ref{3.2}), $s$ is the Riemannian arc
	length and 
	\begin{equation}
		\mathbf{p}(\mathbf{x,\nu })=\nabla \varphi (\mathbf{x,\nu }).  \label{3.50}
	\end{equation}%
	Note that equations (\ref{3.4}), (\ref{3.5}) imply that $\varphi (\mathbf{%
		\xi ,\nu })=s.$ In particular, this means that the second condition (\ref%
	{3.3}) is satisfied. Equations (\ref{3.4}), (\ref{3.5}) define a geodesic
	line $\Gamma (\mathbf{x},\Sigma (\mathbf{\nu }))$ in $D_{+}\left( \mathbf{%
		\nu }\right) $ which connects the point $\mathbf{x}\in D_{+}\left( \mathbf{%
		\nu }\right) $ with the point $\mathbf{y}\in \Sigma (\mathbf{\nu })$ and
	orthogonal to $\Sigma (\mathbf{\nu })$ at $\mathbf{y}$.
	
	Cauchy problem (\ref{3.4}), (\ref{3.5}) has the unique solution 
	\begin{equation}
		\mathbf{x}=\mathbf{f}(s,a_{2},a_{3}),~\mathbf{p}=\mathbf{g}(s,a_{2},a_{3}).
		\label{400}
	\end{equation}%
	These equations define the bundle of geodesic lines, which go out from
	different points $\mathbf{y}\in \Sigma (\mathbf{\nu })$ in direction $%
	\mathbf{\nu }$. To find the geodesic line $\Gamma (\mathbf{x},\Sigma (%
	\mathbf{\nu }))$, we need to invert the first equation \eqref{400} and
	calculate $a_{2}$ and $a_{3}$ and then to find $\mathbf{y}\in \Sigma (%
	\mathbf{\nu })$, using formula \eqref{3.2}. Set 
	\begin{equation*}
		D_{+}(\mathbf{\nu },R)=\{\mathbf{x}:\,-R\leq \mathbf{x}\cdot \mathbf{\nu }%
		\leq R\},\text{ }T=\max_{\mathbf{x}\in D_{+}(\mathbf{\nu },R)}\varphi (%
		\mathbf{x,\nu }).
	\end{equation*}%
	Hence, $D_{+}(\mathbf{\nu },R)\subset D_{+}(\mathbf{\nu }).$ Note that both
	sets $D_{+}(\mathbf{\nu })$ and $D_{+}(\mathbf{\nu },R)$ are closed ones,
	i.e. $D_{+}(\mathbf{\nu })=\overline{D_{+}(\mathbf{\nu })}$ and $D_{+}(%
	\mathbf{\nu },R)=\overline{D_{+}(\mathbf{\nu },R)}.$ Here $T$ is a finite
	number. Indeed, let $C(\mathbf{\nu },R)$ be the circular cylinder with the
	circle of the radius $R$ and the axis $\mathbf{x}=(-R+s^{\prime })\mathbf{%
		\nu }$, $s^{\prime }\geq 0$, with generating lines orthogonal to the plane $%
	\Sigma (\mathbf{\nu }).$ Consider the intersection $D_{0}(\mathbf{\nu },R)$
	of $C(\mathbf{\nu },R)$ with $D_{+}(\mathbf{\nu },R),$ 
	\begin{equation}
		D_{0}(\mathbf{\nu },R)=C(\mathbf{\nu },R)\cap D_{+}(\mathbf{\nu },R).
		\label{2.100}
	\end{equation}%
	Then the function $\varphi (\mathbf{x,\nu })=\mathbf{x}\cdot \mathbf{\nu }%
	+R\leq 2R$ for $\mathbf{x}\in D_{+}(\mathbf{\nu },R)\diagdown D_{0}(\mathbf{%
		\nu },R)$, since $\mathbf{x}\cdot \mathbf{\nu }\in \lbrack -R,R]$ in $D_{+}(%
	\mathbf{\nu },R)$. On the other hand the domain $D_{0}(\mathbf{\nu },R)$ is
	finite. Therefore, 
	\begin{equation*}
		T=\max \left( 2R,\max_{\mathbf{x}\in D_{0}(\mathbf{\nu },R)}\varphi (\mathbf{%
			x,\nu })\right) .
	\end{equation*}
	
	Consider the Jacobian 
	\begin{equation}
		J(\mathbf{x})=\frac{\partial (x_{1},x_{2},x_{3})}{\partial (s,a_{2},a_{3})}.
		\label{2.10}
	\end{equation}%
	From relations \eqref{3.4} and \eqref{3.5} follows that $\partial \mathbf{x}%
	/\partial s=\mathbf{p}=\mathbf{\nu }$ and $\partial \mathbf{x}/\partial
	a_{k}=\mathbf{e}_{k}$, $k=2,3$, at $s=0$. Then the Jacobian is the
	determinant which rows are formed by components of three unite orthogonal
	vectors of the positive orientation. Hence, 
	\begin{equation}
		J(\mathbf{x})=1\text{ for }\mathbf{x}\in \Sigma (\mathbf{\nu }).
		\label{2.1000}
	\end{equation}%
	Note that $J(\mathbf{x})=1$ for $\mathbf{x}\in D_{0}(\mathbf{\nu },R)$ as
	well since $\mathbf{x}=\mathbf{y}+s\mathbf{\nu }$ in $D_{0}(\mathbf{\nu },R)$%
	.
	
	Denote by $c(\mathbf{x})=1/\sqrt{\varepsilon (\mathbf{x})}$ the speed of
	propagation of electromagnetic waves. By (\ref{2.20}) 
	\begin{equation*}
		c_{0}\leq c(\mathbf{x})\leq 1,~\mathbf{x}\in \mathbb{R}^{3},
	\end{equation*}%
	where $c_{0}=1/\sqrt{\varepsilon_0}$.
	
	Below we use the following assumptions:
	
	\medskip \textbf{Assumptions:}
	
	1.\textbf{\ } \emph{The function} $\varepsilon (\mathbf{x})\in C^{\infty }(%
	\mathbb{R}^{3}),$ \emph{satisfies conditions (\ref{2.20}).}
	
	2.\emph{\ There exists a positive constant }$J_{0}$\emph{\ such that} $J(%
	\mathbf{x})\geq J_{0}$ for $D_{+}(\mathbf{\nu },R )$.
	
	3. \emph{Any point }$\mathbf{x}\in D_{+}(\mathbf{\nu },R )$\emph{\ can be
		connected with the plane }$\Sigma \left( \mathbf{\nu }\right) $\emph{\ by a
		single geodesic line }$\Gamma (\mathbf{x},\Sigma (\nu ))$\emph{\ such that }$%
	\Gamma (\mathbf{x},\Sigma (\mathbf{\nu }))$\emph{\ is orthogonal to }$\Sigma
	\left( \mathbf{\nu }\right) $\emph{\ at a point} $\mathbf{y}\in \Sigma
	\left( \mathbf{\nu }\right) $.
	
	4. \emph{Any two points }$\mathbf{x}$ \emph{and} $\mathbf{y}$ \emph{in} $%
	\mathbb{R}^{3}$ \emph{\ can be connected by a single geodesic line.}
	
	\medskip Under these Assumptions, the equality $\mathbf{x}=\mathbf{f}%
	(s,a_{2},a_{3})$ is invertible in $D_{+}(\mathbf{\nu },R)$ and defines $s=s(%
	\mathbf{x},\mathbf{\nu })=\varphi (\mathbf{x,\nu })$ and parameters $%
	a_{k}=a_{k}(\mathbf{x,\nu })$, $k=2,3$, i.e. the point $\mathbf{y}\in \Sigma
	\left( \mathbf{\nu }\right) $, see (\ref{3.2}). By (\ref{3.4}) and (\ref{3.5}%
	) if $\mathbf{x}=\mathbf{y},$ then $d\mathbf{x}/ds=\mathbf{\nu }$. The
	latter vector is directed along the geodesic line $\Gamma (\mathbf{x},\Sigma
	(\mathbf{\nu })).$ Hence, $\Gamma (\mathbf{x},\Sigma (\mathbf{\nu }))$ is
	orthogonal to $\Sigma \left( \mathbf{\nu }\right) $ at the point $\mathbf{y}$%
	.
	
	Define domains $G_{T}\left( \mathbf{\nu }\right) $ and $G\left( T,\mathbf{%
		\nu }\right) $ in $\mathbb{R}^{4}$ as 
	\begin{equation}
		G_{T}\left( \mathbf{\nu }\right) =\{(\mathbf{x},t):\,0\leq t<\min (|\varphi (%
		\mathbf{x,\nu })|,T)\},  \label{500}
	\end{equation}%
	\begin{equation}
		G\left( T,\mathbf{\nu }\right) =\{(\mathbf{x},t)\,,|\varphi (\mathbf{x,\nu }%
		)|\leq t\leq T\}.  \label{501}
	\end{equation}
	
	To differentiate between notations of the Heaviside function $H(t)$ and the
	magnetic wave field $\mathbf{H}(\mathbf{x},t),$ it is convenient to denote $%
	\theta _{0}(t):=$ $H(t),$%
	\begin{equation*}
		\theta _{0}(t)=\left\{ 
		\begin{array}{c}
			1,t\ge 0, \\ 
			0,t<0.%
		\end{array}%
		\right.
	\end{equation*}
	
	\bigskip \textbf{Theorem 1.} \emph{Assume that the Assumption holds. Then
		for every vector} $\mathbf{\nu \in }\mathbb{S}^{2}$ \emph{the solution of
		problem (\ref{2.8}), (\ref{2.9}) can be represented in }$\mathbb{R}%
	_{T}^{4}=\{(\mathbf{x},t)|\,0\le t\leq T\}$ \emph{in the form} 
	\begin{equation}
		\mathbf{E}(\mathbf{x},t)=\mathbf{\alpha }^{-1}(\mathbf{x})\delta (t-\varphi (%
		\mathbf{x}))+\mathbf{\widehat{E}}(\mathbf{x},t)\theta_0 (t-|\varphi (\mathbf{%
			x})|),  \label{10}
	\end{equation}%
	\emph{where} $\mathbf{\alpha }^{-1}(\mathbf{x})\in C^{\infty }(D_{+}(\mathbf{%
		\nu },R ))$, $\mathbf{\alpha }^{-1}(\mathbf{x})=0$ \emph{for} $\mathbf{x}\in
	D_{-}$ \emph{and} $\mathbf{\widehat{E}}(\mathbf{x},t)\in C^{2}\left( G(T,%
	\mathbf{\nu })\right)\,$ \emph{and} $\mathbf{\widehat{E}}(\mathbf{x},t)=0$ 
	\emph{for} $t= |\varphi(\mathbf{x},\mathbf{\nu})|$.
	
	\bigskip \textbf{Proof}. Introduce functions $\theta _{k}(t)$ as%
	\begin{equation}
		\theta _{-3}(t)=\delta ^{\prime \prime }(t),\text{ }\theta _{-2}(t)=\delta
		^{\prime }(t),\text{ }\theta _{-1}(t)=\delta (t),\text{ }\theta _{k}(t)=%
		\frac{t^{k}}{k!}\theta _{0}(t),\quad k=1,2,\ldots .  \label{110}
	\end{equation}%
	Observe that $\theta _{k}^{\prime }(t)=\theta _{k-1}(t)$ for all $k\geq -2$.
	We seek the solution of problem (\ref{2.8}), (\ref{2.9}) in the form 
	\begin{equation}
		\mathbf{E}(\mathbf{x},t)=\sum\limits_{k=-1}^{r}\mathbf{\alpha }^{k}(\mathbf{x%
		})\theta _{k}(t-\varphi (\mathbf{x}))+\mathbf{E}^{r}(\mathbf{x},t),
		\label{11}
	\end{equation}%
	where the natural number $r$ will be chosen later. Substituting
	representation (\ref{11}) in (\ref{2.8}), using the eikonal equation (\ref%
	{3.3}) and equating coefficients at $\theta _{k}(t)$ for $k=-2,-1,0,1,\ldots
	,r-1$, we obtain the following reqursive formulas for finding coefficients $%
	\mathbf{\alpha }^{k}(\mathbf{x})$: 
	\begin{eqnarray}
		2(\nabla \varphi (\mathbf{x})\cdot \nabla )\mathbf{\alpha }^{k}(\mathbf{x})+%
		\mathbf{\alpha }^{k}(\mathbf{x})\Delta \varphi (\mathbf{x})+(\mathbf{\alpha }%
		^{k}(\mathbf{x})\cdot \nabla \ln n^{2}(\mathbf{x}))\nabla \varphi (\mathbf{x}%
		)  \label{12} \\
		=\Delta \mathbf{\alpha }^{k-1}(\mathbf{x})+\nabla (\mathbf{\alpha }^{k-1}(%
		\mathbf{x})\cdot \nabla \ln n^{2}(\mathbf{x})),~k=-1,0,1,...,r.  \notag
	\end{eqnarray}%
	Here we need to formally set 
	\begin{equation}
		\mathbf{\alpha }^{-2}(\mathbf{x})=0.  \label{120}
	\end{equation}%
	Since by (\ref{2.2}) and (\ref{2.9}) $\mathbf{E}=\mathbf{j}\,\delta (t-(%
	\mathbf{x}\cdot \mathbf{\nu }+R))$ for $t<0$, then we obtain 
	\begin{equation}
		\begin{array}{lll}
			\mathbf{\alpha }^{-1}(\mathbf{x})=0,\>\text{in }\>D_{-}; & \mathbf{\alpha }%
			^{-1}|_{\Sigma ({\mathbf{\nu )}}}=\mathbf{j}, &  \\ 
			\mathbf{\alpha }^{k}(\mathbf{x})=0,\>\text{in }\>D_{-}; & \mathbf{\alpha }%
			^{k}|_{\Sigma ({\mathbf{\nu )}}}=0, & k=0,1,\ldots ,r.%
		\end{array}
		\label{13}
	\end{equation}%
	Moreover, using (\ref{2.8}), we obtain the following Cauchy problem for the
	residual $\mathbf{E}^{r}(\mathbf{x},t)$ of expansion (\ref{11}) 
	\begin{equation}
		n^{2}(\mathbf{x})\partial _{t}^{2}\mathbf{E}^{r}-\Delta \mathbf{E}%
		^{r}-\nabla (\mathbf{E}^{r}\cdot \nabla \ln n^{2}(\mathbf{x}))=\mathbf{F}%
		^{r}(\mathbf{x},t),~\mathbf{E}^{r}|_{{t<0}}=0,  \label{14}
	\end{equation}%
	where 
	\begin{equation}
		\mathbf{F}^{r}(\mathbf{x},t)=\left( \Delta \mathbf{\alpha }^{r}(\mathbf{x}%
		)+\nabla \left( \mathbf{\alpha }^{r}(\mathbf{x})\cdot \nabla \ln n^{2}(%
		\mathbf{x})\right) \right) \theta _{r}(t-\varphi (\mathbf{x})).  \label{15}
	\end{equation}%
	We now construct solutions of equation (\ref{12}) with the Cauchy data (\ref%
	{13}). We have along the geodesic line $\Gamma (\mathbf{x},\Sigma (\mathbf{%
		\nu })):$ 
	\begin{equation}
		2(\nabla \varphi (\mathbf{\xi })\cdot \nabla )\mathbf{\alpha }^{k}(\mathbf{%
			\xi })=2(\mathbf{p}(\mathbf{\xi })\cdot \nabla )\mathbf{\alpha }^{k}(\mathbf{%
			\xi })=2n^{2}(\mathbf{\xi })\left( \frac{d\mathbf{\xi }}{ds}\cdot \nabla
		\right) \mathbf{\alpha }^{k}(\mathbf{\xi })  \label{16}
	\end{equation}%
	\begin{equation*}
		=2n^{2}(\mathbf{\xi })\frac{d\mathbf{\alpha }^{k}(\mathbf{\xi })}{ds},
	\end{equation*}%
	where $\mathbf{\xi }\in \Gamma (\mathbf{x},\Sigma (\mathbf{\nu }))$ is an
	arbitrary point. Moreover, it is stated in the paper \cite{R1} that the
	following formula valid along $\Gamma (\mathbf{x},\Sigma (\mathbf{\nu }))$
	(see Lemma 1, Equation (4.2)): 
	\begin{equation}
		\frac{d\ln J(\mathbf{\xi })}{ds}=\mathrm{div}\,\left( n^{-2}(\mathbf{\xi }%
		)\nabla \varphi (\mathbf{\xi })\right) ,  \label{2.11}
	\end{equation}%
	where $J(\mathbf{\xi })$ is the Jacobian defined in (\ref{2.10}), provided
	that $\mathbf{x}=\left( x_{1},x_{2},x_{3}\right) $ is replaced with $\mathbf{%
		\xi }=\left( \xi _{1},\xi _{2},\xi _{3}\right) $. Calculating the
	right-hand-side of (\ref{2.11}) and using (\ref{2.11}), we find%
	\begin{equation*}
		\mathrm{div}\,\left( n^{-2}(\mathbf{\xi })\nabla \varphi (\mathbf{\xi }%
		)\right) =n^{-2}(\mathbf{\xi })\Delta \varphi (\mathbf{\xi })+\left( \nabla
		n^{-2}(\mathbf{\xi })\cdot \mathbf{p}(\mathbf{\xi })\right) .
	\end{equation*}%
	Hence, (\ref{2.11}) implies 
	\begin{equation}
		n^{-2}(\mathbf{\xi })\Delta \varphi (\mathbf{\xi })=\frac{d\ln J(\mathbf{\xi 
			})}{ds}-\left( \nabla n^{-2}(\mathbf{\xi })\cdot \mathbf{p}(\mathbf{\xi }%
		)\right) =\frac{d\ln (J(\mathbf{\xi })n^{2}(\mathbf{\xi }))}{ds}.  \label{17}
	\end{equation}%
	We have used here that, similarly \eqref{16}, 
	\begin{equation*}
		-\left( \nabla n^{-2}(\mathbf{\xi })\cdot p(\mathbf{\xi })\right) = -n^2(%
		\mathbf{\xi })\left( \nabla n^{-2}(\mathbf{\xi })\cdot \frac{d\mathbf{\xi }}{%
			d s}\right) = -n^{2}(\mathbf{\xi })\frac{d}{ds}n^{-2}(\mathbf{\xi })=\frac{d%
		}{ds}\ln n^{2}(\mathbf{\xi }).
	\end{equation*}
	
	Using formulae (\ref{16}) and (\ref{17}) and replacing in (\ref{12}) $%
	\mathbf{x}$ with $\mathbf{\xi }$, we obtain 
	\begin{eqnarray}
		2n^{2}({\mathbf{\xi }})\left[ \frac{d\mathbf{\alpha }^{k}(\mathbf{\xi })}{ds}%
		+\mathbf{\alpha }^{k}(\mathbf{\xi })\frac{d\ln (\sqrt{J(\mathbf{\xi })}n(%
			\mathbf{\xi }))}{ds}\right] +(\mathbf{\alpha }^{k}(\mathbf{\xi })\cdot
		\nabla \ln n^{2}(\mathbf{\xi }))\nabla \varphi (\mathbf{\xi })  \notag \\
		=\Delta \mathbf{\alpha }^{k-1}(\mathbf{\xi })+\nabla (\mathbf{\alpha }^{k-1}(%
		\mathbf{\xi })\cdot \nabla \ln n^{2}(\mathbf{\xi })),~k=-1,0,1,...,r.
		\label{12a}
	\end{eqnarray}%
	Multiplying equation \eqref{12a} by $\sqrt{J(\mathbf{\xi })}/\left( 2n(%
	\mathbf{\xi })\right) $, we transform equation (\ref{12}) along $\Gamma (%
	\mathbf{x},\Sigma (\mathbf{\nu }))$ to the recursive form 
	\begin{eqnarray}
		\frac{d}{ds}\left( \mathbf{\alpha }^{k}(\mathbf{\xi })n(\mathbf{\xi })\sqrt{%
			J(\mathbf{\xi })}\right) -\frac{n(\mathbf{\xi })\sqrt{J(\mathbf{\xi })}}{2}%
		\left( \mathbf{\alpha }^{k}(\mathbf{\xi })\cdot \nabla n^{-2}(\mathbf{\xi }%
		)\right) \mathbf{p}(\mathbf{\xi })  \notag \\
		=\mathbf{Q}^{k}(\mathbf{\xi }),k=-1,0,1,\ldots ,r,  \label{18}
	\end{eqnarray}
	were 
	\begin{equation}
		\mathbf{Q}^{k}(\mathbf{\xi })=\frac{\sqrt{J(\mathbf{\xi })}}{2n(\mathbf{\xi }%
			)}\big[\Delta \mathbf{\alpha }^{k-1}(\mathbf{\xi })+\nabla \left( \mathbf{%
			\alpha }^{k-1}(\mathbf{\xi })\cdot \nabla \ln n^{2}(\mathbf{\xi })\right) %
		\big].  \label{2.12}
	\end{equation}%
	When we solve equations \eqref{18} going \ from $k$ to $k+1$, the function $%
	\mathbf{Q}^{k}(\mathbf{\xi })$ is always known from the previous step. It
	follows from (\ref{120}) and (\ref{2.12}) that $\mathbf{Q}^{-1}(\mathbf{\xi }%
	)=0$. Functions $\mathbf{\alpha }^{k}(\mathbf{\xi })$ satisfy on $\Sigma (%
	\mathbf{\nu })$ conditions (\ref{13}). We also recall that by (\ref{2.20})
	and (\ref{2.100}) $n\left( \mathbf{\xi }\right) |_{\Sigma (\mathbf{\nu })}=1$%
	, $J|_{\Sigma (\mathbf{\nu })}=1.$
	
	Hence, integrating (\ref{18}) with respect to $s$, we obtain a recursive
	integral equation along the geodesic line $\Gamma (\mathbf{x},\Sigma (%
	\mathbf{\nu }))$ 
	\begin{eqnarray}
		\mathbf{\alpha }^{k}(\mathbf{x}) &=&\frac{1}{n(\mathbf{x})\sqrt{J(\mathbf{x})%
		}}\bigg(\mathbf{A}^{k}+\int\limits_{\Gamma (\mathbf{x},\Sigma (\mathbf{\nu }%
			))}\left[ \mathbf{Q}^{k}(\mathbf{\xi })\right.  \notag \\
		&&\left. +\frac{n(\mathbf{\xi })\sqrt{J(\mathbf{\xi })}}{2}\left( \mathbf{%
			\alpha }^{k}(\mathbf{\xi })\cdot \nabla n^{-2}(\mathbf{\xi })\right) p(%
		\mathbf{\xi })\right] \,ds\bigg),~  \label{2.36} \\
		k &=&-1,0,1,\ldots ,r,  \notag
	\end{eqnarray}%
	where $\mathbf{\xi }=\mathbf{f}(s,a_{2},a_{3})$, the vector function $%
	\mathbf{f}(s,a_{2},a_{3})$ is defined in (\ref{400}) and 
	\begin{equation*}
		\mathbf{A}^{-1}=\mathbf{j},~\mathbf{A}^{k}=0,~k=0,1,\ldots ,r.
	\end{equation*}
	
	Recall that $s$ is the Riemannian arc length of $\Gamma (\mathbf{\xi }%
	,\Sigma (\mathbf{\nu }))$. Equation (\ref{2.36}) is a Volterra-type integral
	equation of the second kind along the curve $\Gamma (\mathbf{x},\Sigma (%
	\mathbf{\nu }))$. Therefore, this equation can be solved by the method of
	successive approximations which is rapidly converging. We can solve equation
	(\ref{2.36}) for different $k=-1,...,r$ step-by-step, starting from $k=-1$.
	
	As it will be clear in the sequel, the most important role plays the
	function $\mathbf{\alpha }^{-1}(\mathbf{x})$. We now represent this function
	through the resolvent $R(\mathbf{x},\mathbf{\xi })$ of equation (\ref{2.36})
	for $k=-1$. First, we introduce the vector function $\mathbf{\beta (x)}$ 
	\begin{equation*}
		\mathbf{\beta (x)}={n(\mathbf{x})\sqrt{J(\mathbf{x})}}\mathbf{\alpha }^{-1}(%
		\mathbf{x}).
	\end{equation*}%
	Then equation for this function has the form 
	\begin{equation*}
		\mathbf{\beta (x)}=\mathbf{j}+\frac{1}{2}\int\limits_{\Gamma (\mathbf{x}%
			,\Sigma (\mathbf{\nu }))}\left( \mathbf{\beta (\xi )}\cdot \nabla n^{-2}(%
		\mathbf{\xi })\right) \mathbf{p}(\mathbf{\xi })\,ds.
	\end{equation*}%
	The more convenient form of this equation is: 
	\begin{equation*}
		\mathbf{\beta }(\mathbf{x})=\mathbf{j}+\int\limits_{\Gamma (\mathbf{x}%
			,\Sigma (\mathbf{\nu }))}\mathbf{\beta }(\mathbf{\xi })K_{0}(\mathbf{\xi }%
		)\,ds,
	\end{equation*}%
	where $K_{0}({\mathbf{\xi }})$ is the $3\times 3$ matrix 
	\begin{equation*}
		K_{0}({\mathbf{\xi }})=\frac{1}{2}(\nabla n^{-2}(\mathbf{\xi }))^{\ast }%
		\mathbf{p}(\mathbf{\xi })
	\end{equation*}%
	and $(\nabla n^{-2}(\mathbf{\xi }))^{\ast }$ is the transposed vector $%
	(\nabla n^{-2}(\mathbf{\xi }))$, i.e. column vector, while $\mathbf{p}(%
	\mathbf{\xi })$ is row vector.
	
	Represent $\mathbf{\beta (x)}$ as 
	\begin{equation*}
		\mathbf{\beta (x)}=\sum\limits_{n=0}^{\infty }\mathbf{\beta }^{n}(\mathbf{x}%
		),
	\end{equation*}%
	where 
	\begin{equation*}
		\mathbf{\beta }^{0}(\mathbf{x})=\mathbf{j},~\mathbf{\beta }^{n}(\mathbf{x}%
		)=\int\limits_{\Gamma (\mathbf{x},\Sigma (\mathbf{\nu }))}\mathbf{\beta }%
		^{n-1}(\mathbf{\xi })K_{0}(\mathbf{\xi })\,ds,~n=1,2,\ldots .
	\end{equation*}%
	Then 
	\begin{equation*}
		\mathbf{\beta }^{1}(\mathbf{x})=\mathbf{j}\int\limits_{\Gamma (\mathbf{x}%
			,\Sigma (\mathbf{\nu }))}K_{0}({\mathbf{\xi }})\,ds,
	\end{equation*}%
	Next, 
	\begin{equation*}
		\mathbf{\beta }^{2}(\mathbf{x})=\mathbf{j}\int\limits_{\Gamma (\mathbf{x}%
			,\Sigma (\mathbf{\nu }))}\int\limits_{\Gamma (\mathbf{\xi },\Sigma (\mathbf{%
				\nu }))}K_{0}({\mathbf{\xi }}^{\prime })\,ds^{\prime }K_{0}({\mathbf{\xi }}%
		)\,ds.
	\end{equation*}%
	Changing here the repeated integration by place and then replacing $%
	s^{\prime }$ with $s$ and vice versa, we obtain 
	\begin{equation*}
		\mathbf{\beta }^{2}(\mathbf{x})=\mathbf{j}\int\limits_{\Gamma (\mathbf{x}%
			,\Sigma (\mathbf{\nu }))}K_{1}({\mathbf{x}},{\mathbf{\xi }})\,ds,
	\end{equation*}%
	where 
	\begin{equation*}
		K_{1}(\mathbf{x},\mathbf{\xi })=\int\limits_{\Gamma (\mathbf{x},\Sigma (%
			\mathbf{\nu }))\setminus \Gamma (\mathbf{\xi },\Sigma (\mathbf{\nu }))}K_{0}(%
		\mathbf{\xi })K_{0}(\mathbf{\xi }^{\prime })ds^{\prime }
	\end{equation*}%
	Similarly, 
	\begin{eqnarray*}
		\mathbf{\beta }^{n}(\mathbf{x}) &=&\mathbf{j}\int\limits_{\Gamma (\mathbf{x}%
			,\Sigma (\mathbf{\nu }))}\int\limits_{\Gamma (\mathbf{\xi },\Sigma (\mathbf{%
				\nu }))}K_{n-2}(\mathbf{\ \xi },{\mathbf{\xi }}^{\prime })\,ds^{\prime
		}K_{0}({\mathbf{\xi }})\,ds \\
		&=&\mathbf{j}\int\limits_{\Gamma (\mathbf{x},\Sigma (\mathbf{\nu }))}K_{n-1}(%
		{\mathbf{x}},{\mathbf{\xi }})\,ds,~n\geq 2,
	\end{eqnarray*}%
	where 
	\begin{equation*}
		K_{n-1}(\mathbf{x},\mathbf{\xi })=\int\limits_{\Gamma (\mathbf{x},\Sigma (%
			\mathbf{\nu }))\setminus \Gamma (\mathbf{\xi },\Sigma (\mathbf{\nu }%
			))}K_{n-2}(\mathbf{\xi }^{\prime },\mathbf{\xi })K_{0}(\mathbf{\xi }^{\prime
		})ds^{\prime }
	\end{equation*}%
	and $K_{0}(\mathbf{x},\mathbf{\xi })=K_{0}(\mathbf{\xi })$. Thus, we obtain 
	\begin{equation*}
		\mathbf{\beta (x)}=\mathbf{j}\left( I+\int\limits_{\Gamma (\mathbf{x},\Sigma
			(\mathbf{\nu }))}R(\mathbf{x},\mathbf{\xi })ds\right) .
	\end{equation*}%
	Here $I$ is the identity matrix and $R(\mathbf{x},\mathbf{\xi })$ is defined
	as 
	\begin{equation}
		R(\mathbf{x},\mathbf{\xi })=\sum_{n=0}^{\infty }K_{n}(\mathbf{x},\mathbf{\xi 
		}).  \label{19b}
	\end{equation}%
	Finally we obtain for $\mathbf{\alpha }^{-1}(\mathbf{x})$ the following
	formula 
	\begin{equation*}
		\mathbf{\alpha }^{-1}(\mathbf{x})=\frac{{\mathbf{j}}}{n(\mathbf{x})\sqrt{J(%
				\mathbf{x})}}\left( I+\int\limits_{\Gamma (\mathbf{x},\Sigma (\mathbf{\nu }%
			))}R(\mathbf{x},\mathbf{\xi })ds\right) .
	\end{equation*}%
	Applying the above technique, similar formulae can be easily derived for $%
	\mathbf{\alpha }^{k}(\mathbf{x})$, $k=0,1,\ldots ,r$.
	
	Denote 
	\begin{equation*}
		P^{k}({\mathbf{x}})=\int\limits_{\Gamma (\mathbf{x},\Sigma (\mathbf{\nu }))}%
		\mathbf{Q}^{k}(\mathbf{\xi })\,ds.
	\end{equation*}%
	Then 
	\begin{equation}
		\mathbf{\alpha }^{k}(\mathbf{x})=\frac{1}{n(\mathbf{x})\sqrt{J(\mathbf{x})}}%
		\left( P^{k}({\mathbf{x}})+\int\limits_{\Gamma (\mathbf{x},\Sigma (\mathbf{%
				\nu }))}P^{k}({\mathbf{\xi }})R(\mathbf{x},\mathbf{\xi })ds\right)
		,~k=0,1,\ldots ,r.  \label{20b}
	\end{equation}%
	We estimate $R(\mathbf{x},\mathbf{\xi })$ later in the proof of Theorem 2.
	The uniform convergence of series \eqref{19b} in $D_{+}(\mathbf{x },R )$
	follows from that estimate.
	
	Since the function $n(\mathbf{x})\in C^{\infty }(\mathbb{R}^{3}),$ then
	functions $\mathbf{f}(s,a_{2},a_{3})$ and $J(\mathbf{x})$ belong to $%
	C^{\infty }({D_{+}(\mathbf{\nu },R)})$. Hence, all functions $\mathbf{\alpha 
	}^{k}(\mathbf{x})\in C^{\infty }({D_{+}(\mathbf{\nu },R)})$. Moreover, $%
	\mathbf{\alpha }^{-1}(\mathbf{x})=\mathbf{j}$ and $\mathbf{\alpha }^{k}(%
	\mathbf{x})=0$ for $k=0,1,\ldots ,r$, if $\mathbf{x}$ lies outside $D_{0}(%
	\mathbf{\nu },R)$ since $\text{supp}\nabla \varepsilon (\mathbf{x})\subset B$%
	. Therefore, functions $\mathbf{\alpha }^{k}(\mathbf{x})$ for $k=0,1,\ldots
	,r$ are compactly supported in $D_{+}(\mathbf{\nu },R)$. Hence, $\mathbf{F}%
	^{r}(\mathbf{x},t)=0$ for $\{(\mathbf{x},t)|\,\mathbf{x}\in D_{0}(\mathbf{%
		\nu },R),t\geq 0\}$ (see notation \eqref{2.100}). Moreover, $\mathbf{F}^{r}(%
	\mathbf{x},t)=0$ for $(\mathbf{x},t)\in G_{T}(\mathbf{\nu })$. These two
	facts imply that function $\mathbf{F}^{r}(\mathbf{x},t)$ is compactly
	supported in domain $\mathbb{R}_{T}^{4}\supset G(T,\mathbf{\nu })$. Hence,
	it follows from (\ref{14}) that the vector function $\mathbf{E}^{r}(\mathbf{x%
	},t)$ vanishes in $G_{T}(\mathbf{\nu })$ and it is compactly supported in $%
	\mathbb{R}_{T}^{4}$ because the speed of electromagnetic waves is finite.
	
	We now apply the method of energy estimates to the problem (\ref{14}) in the
	domain $\mathbb{R}_{T}^{4}\supset G(T,\mathbf{\nu })$ to estimate function $%
	\mathbf{E}^{r}(\mathbf{x},t)$. . This method is a powerful tool for
	investigations of various problems of mathematical physics. Applications of
	this methods for boundary value problems are given in \cite{CoHil}, \cite%
	{Lad} and many other books related to partial differential equations. So,
	main ideas of this method are well known. At the same time, we can not give
	a reference of the exact result that we need for our goal. Therefore, for
	the completeness of the proof and for the reader's convenience we formulate
	below a lemma related to the estimate of solution to problem (\ref{14}). Let 
	$Y(t,T)=\mathbb{R}_{T}^{4}\cap \{t=const\}$.
	
	\bigskip \textbf{Lemma.} \emph{Let} $\varepsilon (\mathbf{x})$ \emph{satisfy
		the conditions} 
	\begin{equation}
		\Vert \varepsilon \Vert _{C^{r}(\mathbb{R}^{3})}\leq \mu ,~\Vert \ln
		\varepsilon \Vert _{C^{r+1}(\mathbb{R}^{3})}\leq \mu ,  \label{2.41}
	\end{equation}%
	\emph{with a positive constant} $\mu $ \emph{and function} $\mathbf{F}^{r}(%
	\mathbf{x},t)$ \emph{belongs to} $H^{r}(\mathbb{R}_{T}^{4})$ \emph{and
		satisfies the inequality} 
	\begin{equation}
		\Vert \mathbf{F}^{r}\Vert _{H^{r}(\mathbb{R}_{T}^{4})}\leq M.  \label{2.42}
	\end{equation}%
	\emph{Then the solution of problem} (\ref{14}) $\mathbf{E}^{r}\in
	H^{r+1}(Y(t,T))$ \emph{for all} $t\in \left( 0,T\right) $ \emph{and} \emph{%
		there exists a positive constant} $C_{1}=C_{1}(\mu ,T)$ \emph{such that the
		following estimates hold:}{\ 
		\begin{equation}
			\Vert \mathbf{E}^{r}\Vert _{H^{r+1}(Y(t,T))}\leq C_{1}M,~\Vert \partial _{t}%
			\mathbf{E}^{r}\Vert _{H^{r}(Y(t,T))}\leq C_{1}M.  \label{2.43}
		\end{equation}%
	}
	
	\bigskip The proof of this Lemma is given in the Appendix.
	
	In our case $\mathbf{F}^{r}\in {H^{r}(\mathbb{R}_{T}^{4})}$ and $\varepsilon
	\in C^{\infty }(\mathbb{R}^{3})$. Hence, conditions \eqref{2.41} and %
	\eqref{2.42} valid with some positive $\mu $ and $M$. Applying Lemma, we
	obtain that the solution of problem (\ref{14}) is such that $\mathbf{E}%
	^{r}\in H^{r+1}(Y(t,T))$ and $\partial _{t}\mathbf{E}^{r}\in H^{r}(Y(t,T))$
	for all $t\in \lbrack 0,T]$. Choosing $r=4$ and applying Lemma, we obtain $%
	\mathbf{E}^{4}\in H^{5}(Y(t,T))$ for all $t\in \lbrack 0,T]$ and, hence, $%
	\mathbf{E}^{4}\in H^{5}(\mathbb{R}_{T}^{4})$. Therefore the embedding
	theorem implies that $\mathbf{E}^{4}\in C^{2}(\mathbb{R}_{T}^{4})$. Hence,
	the vector function $\mathbf{E}^{4}\in C^{2}\left( \overline{G(T,\mathbf{\nu 
		})}\right) $ and is continuous together with space derivatives up to the
	second order across the characteristic wedge $t=|\varphi (\mathbf{x})|$. In
	particular, $\mathbf{E}^{4}(\mathbf{x},t)=0$ for $t=|\varphi (\mathbf{x})|$.
	
	Setting 
	\begin{equation}
		\widehat{\mathbf{E}}(\mathbf{x},t)=\sum\limits_{k=0}^{4}\mathbf{\alpha }^{k}(%
		\mathbf{x})\frac{(t-\varphi (\mathbf{x}))^{k}}{k!}+\mathbf{E}^{4}(\mathbf{x}%
		,t),  \label{21}
	\end{equation}%
	we obtain (\ref{10}) as well as the required smoothness $\widehat{\mathbf{E}}%
	\in C^{2}\left( \overline{G(\mathbf{\nu },T)}\right) $. $\square $\textbf{\ }
	
	\bigskip \textbf{Remark 1. }The equality (\ref{21}) implies the following
	formula, which we use below: 
	\begin{equation}
		\lim_{t\rightarrow \varphi (\mathbf{x})^{+}}\widehat{\mathbf{E}}(\mathbf{x}%
		,t)=\mathbf{\alpha }^{0}(\mathbf{x}).  \label{22}
	\end{equation}
	
	\section{Frequency domain}
	
	\label{sec:3}
	
	Consider the Fourier transform $\widetilde{\mathbf{E}}(\mathbf{x},k)$ of the
	function $\mathbf{E}(\mathbf{x},t)$, 
	\begin{eqnarray}
		\widetilde{\mathbf{E}}(\mathbf{x},k) &=&\int\limits_{-\infty }^{\infty }%
		\mathbf{E}(\mathbf{x},t)\exp \left( -ikt\right) dt  \notag \\
		&=&\widetilde{\mathbf{E}}^{0}(\mathbf{x},k)+\int\limits_{0}^{\infty }\mathbf{%
			E}(\mathbf{x},t)\exp \left( -ikt\right) dt,  \label{220}
	\end{eqnarray}%
	where $k=2\pi /\lambda $ is the wave number, $\lambda $ is the dimensionless
	wavelength and 
	\begin{equation*}
		\widetilde{\mathbf{E}}^{0}(\mathbf{x},k)=\mathbf{j}\exp (i(\mathbf{x}\cdot 
		\mathbf{\nu }+R))\theta _{0}(-\mathbf{x}\cdot \mathbf{\nu }-R)
	\end{equation*}%
	is the Fourier transform of $\mathbf{E}^{0}(\mathbf{x},t)$. The existence of
	the integral in (\ref{220}) follows from results of Vainberg \cite{V} which
	claim that the vector function $\mathbf{E}(\mathbf{x},t)$ decays
	exponentially together with its appropriate derivatives as $t\rightarrow
	\infty $ while $\mathbf{x}$ runs over any bounded domain $\Omega \subset 
	\mathbb{R}^{3}$. Next, theorem 3.3 of \cite{V1} and theorem 6 of Chapter 9
	of \cite{V} guarantee that $\widetilde{\mathbf{E}}(\mathbf{x},k)$ is the
	solution to the equation 
	\begin{equation}
		(\Delta +k^{2}n^{2}(\mathbf{x}))\widetilde{\mathbf{E}}+\nabla (\widetilde{%
			\mathbf{E}}\cdot \nabla \ln n^{2}(\mathbf{x}))=0,~\mathbf{x}\in \mathbb{R}%
		^{3},  \label{23}
	\end{equation}%
	where the scattering field 
	\begin{equation*}
		\widetilde{\mathbf{E}}^{sc}(\mathbf{x},k)=\widetilde{\mathbf{E}}(\mathbf{x}%
		,k)-\widetilde{\mathbf{E}}^{0}(\mathbf{x},k)
	\end{equation*}%
	satisfies the radiation condition as $|\mathbf{x}|\rightarrow \infty $.
	
	\bigskip We now consider the vector function $\widetilde{\mathbf{E}}(\mathbf{%
		x},k)$ in \eqref{220} for $\mathbf{x}\in D_{+}(\mathbf{\nu },R )$. Using
	representation (\ref{10}), we obtain 
	\begin{equation}
		\widetilde{\mathbf{E}}(\mathbf{x},k)=\mathbf{\alpha }^{-1}(\mathbf{x})\exp
		(-ik\varphi (\mathbf{x}))+\int\limits_{\varphi (\mathbf{x)}}^{\infty }%
		\widehat{\mathbf{E}}\left( \mathbf{x},t\right) \exp \left( -ikt\right) \,dt.
		\label{24}
	\end{equation}%
	Integrating by parts in (\ref{24}) and using formula, we obtain 
	\begin{eqnarray*}
		\widetilde{\mathbf{E}}(\mathbf{x},k) &=&\mathbf{\alpha }^{-1}(\mathbf{x}%
		)\exp (-ik\varphi (\mathbf{x}))+\frac{\exp (-ik\varphi (\mathbf{x}))}{ik}%
		\mathbf{\alpha }^{0}(\mathbf{x}) \\
		&&+\frac{1}{ik}\int\limits_{\varphi (\mathbf{x})}^{\infty }\widehat{\mathbf{E%
		}}_{t}\left( \mathbf{x},t\right) \exp \left( -ikt\right) dt,\quad
	\end{eqnarray*}%
	Thus,%
	\begin{equation}
		\widetilde{\mathbf{E}}(\mathbf{x},k)=\mathbf{\alpha }^{-1}(\mathbf{x})\exp
		(-ik\varphi (\mathbf{x}))+O\left( \frac{1}{k}\right) ,\quad k\rightarrow
		\infty ,~\forall \,\mathbf{x}\in D_{+}(\mathbf{\nu },R ).  \label{25}
	\end{equation}
	Consider now the equation 
	\begin{equation}
		(\Delta +k^{2}n^{2}(\mathbf{x})\overline{\mathbf{E}}=0,~\mathbf{x}\in 
		\mathbb{R}^{3},  \label{27}
	\end{equation}%
	with the incident plane wave $\overline{\mathbf{E}}^{0}(\mathbf{x},k)=%
	\mathbf{j}\exp (ik(\mathbf{x}\cdot \mathbf{\nu }+R))\theta _{0}(-\mathbf{x}%
	\cdot \mathbf{\nu }-R)$ and the radiation condition for $\left( \overline{%
		\mathbf{E}}-\overline{\mathbf{E}}^{0}\right) $. Note that $\overline{\mathbf{%
			E}}\cdot \mathbf{\nu }=0$ and $\overline{\mathbf{E}}\cdot (\mathbf{\nu }%
	\times \mathbf{j})=0$ since $\overline{\mathbf{E}}^{0}\cdot \mathbf{\nu }=0$
	and $\overline{\mathbf{E}}^{0}\cdot (\mathbf{\nu }\times \mathbf{j})=0$
	because $\overline{\mathbf{E}}^{0}$ is parallel to $\mathbf{j}$ and also $%
	\mathbf{j}\cdot \mathbf{\nu }=0$.
	
	Consider the function $\overline{u}(\mathbf{x},t)=\overline{\mathbf{E}}\cdot 
	\mathbf{j.}$ Then 
	\begin{equation}
		(\Delta +k^{2}n^{2}(\mathbf{x})\overline{u}(\mathbf{x},k)=0,~\mathbf{x}\in 
		\mathbb{R}^{3},  \label{27a}
	\end{equation}%
	with the incident plane wave $\overline{u}^{0}(\mathbf{x},k)=\exp (ik(%
	\mathbf{x}\cdot \mathbf{\nu }+R))\theta _{0}(-\mathbf{x}\cdot \mathbf{\nu }%
	-R)$ and the radiation condition for $\left( \overline{u}-\overline{u}%
	^{0}\right) $.
	
	We impose conditions below, which guarantee that the electric wave field $%
	\overline{\mathbf{E}}_{\mathbf{j}}=\mathbf{j}\overline{u}(\mathbf{x},k)$ is
	close to $\widetilde{\mathbf{E}}$ at the high values of the wave number $k$,
	which is equivalent to small wavelengths $\lambda $. First, suppose that
	these two electric wave fields are indeed close to each other in the norm of
	the space $C\left( \overline{{D_{+}(\mathbf{\nu },R)}}\right) $. This means
	that 
	\begin{equation}
		\widetilde{\mathbf{E}}(\mathbf{x},k)=\overline{\mathbf{E}}(\mathbf{x},k)+%
		\mathbf{V}(\mathbf{x},k),\text{ }\mathbf{x}\in {D_{+}(\mathbf{\nu },R)},
		\label{4.1a}
	\end{equation}%
	\begin{equation}
		\left\Vert \mathbf{V}(\mathbf{x},k)\right\Vert _{C\left( {D_{+}(\mathbf{\nu }%
				,R)}\right) }\leq \sigma ,  \label{4.2a}
	\end{equation}%
	where $\sigma >0$ is a small number. Consider the component $\widetilde{u}(%
	\mathbf{x},k)=\widetilde{\mathbf{E}}(\mathbf{x},k)\cdot \mathbf{j}$. Then%
	\begin{equation}
		\widetilde{u}(\mathbf{x},k)=\overline{u}(\mathbf{x},k)+\mathbf{V}(\mathbf{x}%
		,k)\cdot \mathbf{j,}  \label{4.3a}
	\end{equation}%
	\begin{equation}
		\left\Vert \mathbf{V}(\mathbf{x},k)\cdot \mathbf{j}\right\Vert _{C\left( {%
				D_{+}(\mathbf{\nu },R)}\right) }\leq \sigma ,  \label{4.4a}
	\end{equation}%
	which means that the function $\widetilde{u}(\mathbf{x},k)$ approximates
	well the solution of the Helmholtz equation (\ref{27a}) with the above
	incident plane wave and corresponding radiation conditions.
	
	Next, since\textbf{\ }$\mathbf{j\cdot \nu }=\mathbf{j}\cdot (\mathbf{\nu }%
	\times \mathbf{j})=0,$ then (\ref{4.1}) implies that 
	\begin{equation*}
		\widetilde{\mathbf{E}}(\mathbf{x},k)\cdot \mathbf{\nu =V}(\mathbf{x},k)\cdot 
		\mathbf{\nu ,}
	\end{equation*}%
	\begin{equation*}
		\widetilde{\mathbf{E}}(\mathbf{x},k)\cdot (\mathbf{\nu }\times \mathbf{j})=%
		\mathbf{V}(\mathbf{x},k)\cdot (\mathbf{\nu }\times \mathbf{j}).
	\end{equation*}%
	Hence, components $\widetilde{\mathbf{E}}(\mathbf{x},k)\cdot \mathbf{\nu }$
	and $\widetilde{\mathbf{E}}(\mathbf{x},k)\cdot (\mathbf{\nu }\times \mathbf{j%
	})$ are sufficiently small, 
	\begin{equation}
		\left\Vert \widetilde{\mathbf{E}}(\mathbf{x},k)\cdot \mathbf{\nu }%
		\right\Vert _{C\left( {D_{+}(\mathbf{\nu },R)}\right) }\leq \sigma
		,\left\Vert \widetilde{\mathbf{E}}(\mathbf{x},k)\cdot \mathbf{\nu }%
		\right\Vert _{C\left( {D_{+}(\mathbf{\nu },R)}\right) }\leq \sigma .
		\label{4.5a}
	\end{equation}%
	Thus, it follows from (\ref{4.1a})-(\ref{4.5a}) that the component $%
	\widetilde{u}(\mathbf{x},k)=\widetilde{\mathbf{E}}(\mathbf{x},k)\cdot 
	\mathbf{j}$ \ of the electric wave field dominates two other components and
	it is close to the solution $\overline{u}(\mathbf{x},k)$ of the Helmholtz
	equation supplied by the above incident plane wave and radiation conditions.
	
	\textbf{Remark 2}. Since experimental data have noise, then it is sufficient
	to obtain a good approximation $\overline{u}(\mathbf{x},k)$ for the
	component $\widetilde{u}(\mathbf{x},k)=\widetilde{\mathbf{E}}(\mathbf{x}%
	,k)\cdot \mathbf{j}$ of the electric wave field in the $C\left( \overline{{%
			D_{+}(\mathbf{\nu },R)}}\right) -$norm.
	
	What is left to do is to prove (\ref{4.1a}), (\ref{4.2a}). And this is what
	is done in the rest of this section.
	
	It is easy to derive a complete analog of formula (\ref{25}) for the
	function $\overline{\mathbf{E}}$. To do this, one should consider the Cauchy
	problem for the time dependent analog of (\ref{27}) and repeat arguments of
	Theorem 1 for the case when the term $\nabla (\mathbf{E}\cdot \nabla \ln
	n^{2}(\mathbf{x}))$ is neglected in (\ref{2.8}). Hence, 
	\begin{equation}
		\overline{\mathbf{E}}(\mathbf{x},k)=\widehat{\mathbf{\alpha }}^{-1}(\mathbf{x%
		})\exp (-ik\varphi (\mathbf{x}))+O\left( \frac{1}{k}\right) ,\quad
		k\rightarrow \infty ,~\forall \,\mathbf{x}\in D_{+}(\mathbf{\nu },R ),
		\label{2.52}
	\end{equation}%
	where 
	\begin{equation}
		\widehat{\mathbf{\alpha }}^{-1}(\mathbf{x})=\frac{\mathbf{j}}{n(\mathbf{x})%
			\sqrt{J(\mathbf{x})}}.  \label{2.53}
	\end{equation}
	
	\bigskip \textbf{Theorem 2.} \emph{Suppose that the Assumptions hold. Let }$%
	\eta >0$\emph{\ be such a constant that} 
	\begin{equation}
		\left\Vert \nabla n(\mathbf{x})\right\Vert _{C\left( \overline{B}\right)
		}\leq \eta .  \label{2.60}
	\end{equation}%
	\emph{Then} 
	\begin{equation*}
		\widetilde{\mathbf{E}}(\mathbf{x},k)-\overline{\mathbf{E}}(\mathbf{x},k)=%
		\mathbf{V}(\mathbf{x},k)+O\left( \frac{1}{k}\right) ,~k\rightarrow \infty ,~%
		\mathbf{x}\in D_{+}(\mathbf{\nu },R)
	\end{equation*}%
	\emph{and} 
	\begin{equation}
		\begin{array}{ll}
			\displaystyle\widetilde{\mathbf{E}}(\mathbf{x},k)\cdot \mathbf{j}=\frac{\exp
				(-ik\varphi (\mathbf{x}))}{n(\mathbf{x})\sqrt{J(\mathbf{x})}}+\mathbf{V}(%
			\mathbf{x},k)\cdot \mathbf{j}+O\left( \frac{1}{k}\right) , &  \\ 
			\displaystyle\widetilde{\mathbf{E}}(\mathbf{x},k)\cdot \mathbf{\nu }=\mathbf{%
				V}(\mathbf{x},k)\cdot \mathbf{\nu }+O\left( \frac{1}{k}\right) , &  \\ 
			\displaystyle\widetilde{\mathbf{E}}(\mathbf{x},k)\cdot (\mathbf{\nu }\times 
			\mathbf{j})=\mathbf{V}(\mathbf{x},k)\cdot (\mathbf{\nu }\times \mathbf{j}%
			)+O\left( \frac{1}{k}\right) , & k\rightarrow \infty ,~\mathbf{x}\in D_{+}(%
			\mathbf{\nu },R),%
		\end{array}
		\label{2.54}
	\end{equation}%
	\emph{where} 
	\begin{equation}
		\begin{array}{ll}
			\mathbf{V}(\mathbf{x},k)=(\mathbf{\alpha }^{-1}(\mathbf{x})-\widehat{\mathbf{%
					\alpha }}^{-1}(\mathbf{x}))\exp (-ik(\mathbf{x}\cdot \mathbf{\nu })), &  \\ 
			\displaystyle|\mathbf{V}(\mathbf{x},k)|\leq \frac{1}{\sqrt{J_{0}}}[\exp
			(\eta T)-1], & \mathbf{x}\in D_{+}(\mathbf{\nu },R).%
		\end{array}
		\label{2.55}
	\end{equation}%
	\emph{Thus, if the number }$\eta $\emph{\ in (\ref{2.60}) is sufficiently
		small, then (\ref{4.1a}) and (\ref{4.2a}) hold.}
	
	\bigskip \textbf{Proof.} Using formulae \eqref{10} and \eqref{25} we obtain
	the first relation \eqref{2.55}. Formulae \eqref{2.54} follow from %
	\eqref{2.55} as well as from \eqref{25}. Then, using \eqref{25} and %
	\eqref{2.53}, we obtain 
	\begin{equation*}
		|\mathbf{V}(\mathbf{x},k)|=|\mathbf{\alpha }^{-1}(\mathbf{x})-\widehat{%
			\mathbf{\alpha }}^{-1}(\mathbf{x})|=\frac{1}{n(\mathbf{x})\sqrt{J(\mathbf{x})%
		}}\left\vert \,\int\limits_{\Gamma (\mathbf{x},\Sigma )}R(\mathbf{x},\mathbf{%
			\xi })ds\right\vert .
	\end{equation*}
	
	We now estimate $R(\mathbf{x},\mathbf{\xi })$ for $\mathbf{x}\in D_{+}(%
	\mathbf{\nu },R )$ and $\mathbf{\xi }\in \Gamma (\mathbf{x},\Sigma (\mathbf{%
		\nu }))$. Note that $R(\mathbf{x},\mathbf{\xi })=0$ for $\mathbf{x}\in D_{+}(%
	\mathbf{\nu },R )\setminus D_{0}(\mathbf{\nu },R )$ since $\nabla n(\mathbf{x%
	})=0$ and, hence, all $K_{n}(\mathbf{x},\mathbf{\xi })=0$ in this domain.
	So, we need estimate $R(\mathbf{x},\mathbf{\xi })$ for $\mathbf{x}\in D_{0}(%
	\mathbf{\nu },R )$ only.
	
	Introduce the matrix norm for a matrix $K(\mathbf{x},\mathbf{\xi })=(k_{ij}(%
	\mathbf{x},\mathbf{\xi })_{i,j=1}^{3}$ as 
	\begin{equation*}
		\Vert K(\mathbf{x},\mathbf{\xi }\Vert =\max_{i,j=1,2,3}|k_{ij}(\mathbf{x},%
		\mathbf{\xi })|.
	\end{equation*}%
	Let $\mathbf{x}\in D_{0}(\mathbf{\nu },R)$ and $\varphi (\mathbf{x})=s_{0}$
	and $\varphi (\mathbf{\xi })=s$. Then using \eqref{19b} and formulae for $%
	K_{n}(\mathbf{x},\mathbf{\xi })$, $n=0,1,2,\ldots$, we obtain 
	\begin{equation*}
		\begin{array}{llll}
			\Vert K_{0}(\mathbf{x},\mathbf{\xi })\Vert \leq \eta , & \mathbf{x}\in D_{0}(%
			\mathbf{\nu },R), & \mathbf{\xi }\in \Gamma (\mathbf{x},\Sigma (\mathbf{\nu }%
			)), &  \\ 
			\Vert K_{1}(\mathbf{x},\mathbf{\xi })\Vert \leq \eta ^{2}(s_{0}-s), & 
			\mathbf{x}\in D_{0}(\mathbf{\nu },R), & \mathbf{\xi }\in \Gamma (\mathbf{x}%
			,\Sigma (\mathbf{\nu })), &  \\ 
			\displaystyle\Vert K_{n}(\mathbf{x},\mathbf{\xi })\Vert \leq \eta ^{n+1}%
			\frac{(s_{0}-s)^{n}}{n!}, & \mathbf{x}\in D_{0}(\mathbf{\nu },R), & \mathbf{%
				\xi }\in \Gamma (\mathbf{x},\Sigma (\mathbf{\nu })), & n=2,3,\ldots%
		\end{array}%
	\end{equation*}%
	Hence 
	\begin{equation*}
		\Vert R(\mathbf{x},\mathbf{\xi })\Vert \leq \eta \exp (\eta (s_{0}-s)),~%
		\mathbf{x}\in D_{0}(\mathbf{\nu },R),
	\end{equation*}
	
	Since $n^{-1}(\mathbf{x})\leq 1$ and $J(\mathbf{x})\geq J_{0}$, then we
	arrive at the estimate: 
	\begin{equation*}
		|\mathbf{V}(\mathbf{x},k)|\leq \frac{1}{\sqrt{J_{0}}}[\exp (\eta
		s_{0})-1]\leq \frac{1}{\sqrt{J_{0}}}[\exp (\eta T)-1],~\mathbf{x}\in D_{0}(%
		\mathbf{\nu },R).
	\end{equation*}
	
	Thus, we obtain the estimate in the second line of (\ref{2.55})\emph{. }This
	estimate concludes the proof. $\square $
	
	\section{Relevance to Experimental Results of \protect\cite%
		{Khoa2,Khoa3,KlibKol4}}
	
	\label{sec:4}
	
	We now explain why Theorem 2 at least partially justifies the validity of
	modeling of the propagation of electromagnetic waves in the frequency domain
	by the single Helmholtz equation (\ref{27a}) in the works of the second
	author with coauthors on experimental data \cite{Khoa2,Khoa3,KlibKol4}. We
	say \textquotedblleft at least partially" because a completely precise
	explanation is unlikely possible since we deal here with a sort of a
	\textquotedblleft mathematics-to-physics bridge".
	
	We recall that accurate reconstruction results were obtained in \cite%
	{Khoa2,Khoa3,KlibKol4} when solving coefficient inverse problems.
	Experimental data in these references were collected for the cases when
	rather small inclusions mimicking land mines and improvised explosive
	devices were embedded in an otherwise uniform background (dry sand). The
	dielectric constant was not changing within such an inclusion, although this
	was not an assumption in reconstruction algorithms. Therefore,%
	\begin{equation*}
		\nabla n(\mathbf{x})=\left\{ 
		\begin{array}{c}
			0\text{ within such an inclusion,} \\ 
			0\text{ in the background.}%
		\end{array}%
		\right. 
	\end{equation*}
	The question remains now about the discontinuity of the function $n(\mathbf{x%
	})$ at the inclusion/background interface. Since any solution of an elliptic
	equation, such as, e.g. (\ref{27a}), is sufficiently smooth outside of
	discontinuities of its coefficients \cite{GT}, then we conjecture that the
	medium \textquotedblleft percepts" the functions $n(\mathbf{x})$ in those
	inclusions as a smooth function with rather non-small values of $\left\vert
	\nabla n(\mathbf{x})\right\vert $ near those interfaces. In fact, this has
	been observed in computed images of \cite{Khoa2,Khoa3,KlibKol4}. Now, since
	values of $\left\vert \nabla n(\mathbf{x})\right\vert $ were not small only
	in close proximities of those interfaces and volumes of those proximities
	were small, then this means that norms $\left\Vert \left\vert \nabla n(%
	\mathbf{x})\right\vert \right\Vert _{L_{2}\left( B\right) }$ were actually
	small. On the other hand, since finite differences with relatively small
	numbers of grid points were used in \cite{Khoa2,Khoa3,KlibKol4} to solve
	inverse problems and since all norms in a finite dimensional space are
	equivalent, then the smallness of the discrete norm $\left\Vert \left\vert
	\nabla n(\mathbf{x})\right\vert \right\Vert _{L_{2}\left( B\right) }$ is
	equivalent to the smallness of the discrete norm $\left\Vert \left\vert
	\nabla n(\mathbf{x})\right\vert \right\Vert _{C\left( \overline{B}\right) },$
	which is close to the smallness assumption imposed in Theorem 2 on the
	number $\eta $ in (\ref{2.60}). Note that smallness assumptions were not
	used in algorithms of \cite{Khoa2,Khoa3,KlibKol4}.
	
	Thus, Theorem 2 explains, at least partially, accurate reconstructions in 
	\cite{Khoa2,Khoa3,KlibKol4}.
	
	\section{ Appendix}
	
	\label{sec:5}
	
	\bigskip \textbf{Proof of the Lemma.}
	
	Denote ${E}_{j}^{r}$, $j=1,2,3$, components of vector function $\mathbf{E}%
	^{r}$. Calculating scalar product of both sides of equation \eqref{14}, $%
	2\partial _{t}\mathbf{E}^{r}$, using $n^{2}(\mathbf{x})=\varepsilon (\mathbf{%
		x})$ and applying the identity 
	\begin{equation*}
		2\partial _{t}\mathbf{E}^{r}\cdot \Delta \mathbf{E}^{r}=2\mathrm{div}\left(
		\sum\limits_{j=1}^{3}(\partial _{t}{E}_{j}^{r})\nabla {E}_{j}^{r}\right)
		-\partial _{t}\left( \sum\limits_{j=1}^{3}|\nabla {E}_{j}^{r}|^{2}\right) ,
	\end{equation*}%
	we obtain: 
	\begin{eqnarray}
		\partial _{t}\left( \varepsilon (\mathbf{x})|\partial _{t}\mathbf{E}%
		^{r}|^{2}+\sum\limits_{j=1}^{3}|\nabla {E}_{j}^{r}|^{2}\right) -2\mathrm{div}%
		\left( \sum\limits_{j=1}^{3}(\partial _{t}{E}_{j}^{r})\nabla {E}%
		_{j}^{r}\right)  \notag \\
		-2\sum\limits_{i,j=1}^{3}\partial _{t}{E}_{i}^{r}\left[ (\partial _{x_{i}}{E}%
		_{j}^{r})\partial _{x_{j}}\varepsilon (\mathbf{x})+{E}_{j}^{r}\partial
		_{x_{j}x_{i}}\ln \varepsilon (\mathbf{x})\right] &=&2(\partial _{t}\mathbf{E}%
		^{r})\cdot \mathbf{F}^{r}.  \label{4.1}
	\end{eqnarray}%
	Integrating identity \eqref{4.1} over the domain $\mathbb{R}_{t}^{4}$, $t\in
	(0,T]$ and taking into account that $\mathbf{E}^{r}$ and $\mathbf{F}^{r}$
	are compactly supported in $\mathbb{R}_{T}^{4}$ and the initial zero data,
	we arrive to the equality 
	\begin{eqnarray}
		&&\int\limits_{Y(t,T)}\left( \varepsilon (\mathbf{x})|\partial _{t}\mathbf{E}%
		^{r}(\mathbf{x},t)|^{2}+\sum\limits_{j=1}^{3}|\nabla {E}_{j}^{r}(\mathbf{x}%
		,t)|^{2}\right) d\mathbf{x}  \notag \\
		&=&2\int\limits_{\mathbb{R}_{t}^{4}}\sum\limits_{i,j=1}^{3}\left( \partial
		_{\tau }{E}_{i}^{r}(\mathbf{x},\tau )\right) \left( \partial _{\xi _{i}}{E}%
		_{j}^{r}(\mathbf{x},\tau )\right) \partial _{\xi _{j}}\ln \varepsilon (%
		\mathbf{x})\,d\mathbf{x}d\tau  \notag \\
		&&+2\int\limits_{\mathbb{R}_{t}^{4}}\sum\limits_{i,j=1}^{3}\left( \partial
		_{\tau }{E}_{i}^{r}(\mathbf{x},\tau )\right) \left( {E}_{j}^{r}(\mathbf{x}%
		,\tau )\right) \partial _{\xi _{i}\xi _{j}}\ln \varepsilon (\mathbf{x})\,d%
		\mathbf{x}d\tau  \notag \\
		&&+2\int\limits_{\mathbb{R}_{t}^{4}}\partial _{\tau }\mathbf{E}^{r}(\mathbf{x%
		},\tau )\cdot \mathbf{F}^{r}(\mathbf{x},\tau )\,d\mathbf{x}d\tau .
		\label{4.3}
	\end{eqnarray}%
	Transform this equality using assumption \eqref{2.20}, \eqref{2.41}, the
	algebraic inequalities $2\mathbf{a\cdot b}\leq |\mathbf{a}|^{2}+|\mathbf{b}%
	|^{2}$, we obtain 
	\begin{eqnarray}
		&&\varepsilon _{0}\int\limits_{Y(t,T)}\left( |\partial _{t}\mathbf{E}^{r}(%
		\mathbf{x},t)|^{2}+\sum\limits_{j=1}^{3}|\nabla {E}_{j}^{r}(\mathbf{x}%
		,t)|^{2}\right) d\mathbf{x}  \notag \\
		&\leq &C_{1}\int\limits_{\mathbb{R}_{t}^{4}}\left[ |\partial _{\tau }\mathbf{%
			E}^{r}(\mathbf{x},\tau )|^{2}+\sum\limits_{j=1}^{3}|\nabla {E}_{j}^{r}(%
		\mathbf{x},\tau )|^{2}+|\mathbf{E}^{r}(\mathbf{x},\tau )|^{2}\right] \,d%
		\mathbf{x}d\tau  \notag \\
		&&+\int\limits_{\mathbb{R}_{t}^{4}}|\mathbf{F}^{r}(\mathbf{x},\tau )\,d%
		\mathbf{x}|^{2}\,d\tau .  \label{4.5}
	\end{eqnarray}%
	Using the inequality 
	\begin{equation*}
		|\mathbf{F}^{r}(\mathbf{x},t)|^{2}=\left( \int\limits_{0}^{t}\partial _{\tau
		}\mathbf{E}^{r}(\mathbf{x},\tau )\,d\tau \right) ^{2}\leq
		T\int\limits_{0}^{t}|\partial _{\tau }\mathbf{E}^{r}(\mathbf{x},\tau
		)|^{2}\,d\tau ,
	\end{equation*}%
	we obtain the more general inequality 
	\begin{eqnarray}
		&&\int\limits_{Y(t,T)}\left( |\partial _{t}\mathbf{E}^{r}(\mathbf{x}%
		,t)|^{2}+\sum\limits_{j=1}^{3}|\nabla {E}_{j}^{r}(\mathbf{x},t)|^{2}+|%
		\mathbf{E}^{r}(\mathbf{x},t)|^{2}\right) d\mathbf{x}  \notag \\
		&\leq &C_{1}\int\limits_{\mathbb{R}_{t}^{4}}\left[ |\partial _{\tau }\mathbf{%
			E}^{r}(\mathbf{x},\tau )|^{2}+\sum\limits_{j=1}^{3}|\nabla {E}_{j}^{r}(%
		\mathbf{x},\tau )|^{2}+|\mathbf{E}^{r}(\mathbf{x},\tau )|^{2}\right] \,d%
		\mathbf{x}d\tau  \notag \\
		&&+\int\limits_{\mathbb{R}_{t}^{4}}|\mathbf{F}^{r}(\mathbf{x},\tau )|^{2}\,d%
		\mathbf{x}d\tau .  \label{4.6}
	\end{eqnarray}
	
	Applying Gronwall-Bellman to inequality \eqref{4.6}, we find 
	\begin{equation*}
		\int\limits_{Y(t,T)}\left( |\partial _{t}\mathbf{E}^{r}(\mathbf{x}%
		,t)|^{2}+\sum\limits_{j=1}^{3}|\nabla {E}_{j}^{r}(\mathbf{x},t)|^{2}+|%
		\mathbf{E}^{r}(\mathbf{x},t)|^{2}\right) d\mathbf{x}\leq \Vert \mathbf{F}%
		^{r}\Vert _{\mathbb{R}_{T}^{4}}^{2}\exp (C_{1}T).
	\end{equation*}
	
	Thus, we have obtained the inequalities: 
	\begin{equation}
		\Vert \mathbf{E}^{r}\Vert _{H^{1}(Y(t,T))}\leq C_{1}M,~\Vert \partial _{t}%
		\mathbf{E}^{r}\Vert _{L^{2}(Y(t,T))}\leq C_{1}M.  \label{4.8}
	\end{equation}%
	where $M$ is defined in \eqref{2.42}.
	
	Differentiating equation \eqref{14} $k\leq r$ times with respect to $t$ and
	then calculating scalar product of both sides of the resulting equation with
	the vector function $2\partial _{t}^{k}\mathbf{E}^{r}$, we obtain relations %
	\eqref{4.1}-\eqref{4.8} with $\partial _{t}^{k}\mathbf{E}^{r}$ instead $%
	\mathbf{E}^{r}$. Therefore, the following estimates hold 
	\begin{equation}
		\Vert \partial _{t}^{k}\mathbf{E}^{r}\Vert _{H^{1}(Y(t,T))}\leq
		C_{1}M,~\Vert \partial _{t}^{k+1}\mathbf{E}^{r}\Vert _{L^{2}(Y(t,T))}\leq
		C_{1}M,~k\leq r.  \label{4.9}
	\end{equation}
	
	Apply now the mathematical induction method to prove estimate \eqref{2.43}.
	Suppose that for some $n$, $1<n-1<r-1$, the estimates similar \eqref{4.8}, %
	\eqref{4.9} hold: 
	\begin{equation}
		\begin{array}{ll}
			\Vert \mathbf{E}^{r}\Vert _{H^{n-1}(Y(t,T))}\leq C_{1}M,~\Vert \partial _{t}%
			\mathbf{E}^{r}\Vert _{H^{n-2}(Y(t,T))}\leq C_{1}M, &  \\ 
			\Vert \partial _{t}^{k}\mathbf{E}^{r}\Vert _{H^{n-1}(Y(t,T))}\leq
			C_{1}M,~\Vert \partial _{t}^{k+1}\mathbf{E}^{r}\Vert _{H^{n-2}(Y(t,T))}\leq
			C_{1}M, & k\leq r-(n-2),%
		\end{array}
		\label{4.10}
	\end{equation}%
	and prove that the similar estimates are valid when $n-1$ replaced with $n$.
	Denote 
	\begin{equation*}
		D^{\alpha }=\frac{\partial ^{|\alpha |}}{\partial _{x_{1}}^{\alpha
				_{1}}\partial _{x_{2}}^{\alpha _{2}}\partial _{x_{3}}^{\alpha _{3}}},
	\end{equation*}%
	where $\alpha =(\alpha _{1},\alpha _{2},\alpha _{3})$ is a multi index, $%
	\alpha _{1},\alpha _{2},\alpha _{3}$ are integer nonnegative numbers and $%
	|\alpha |=\alpha _{1}+\alpha _{2}+\alpha _{3}$. We shall use the Leibnitz
	formula for a product of two functions 
	\begin{equation*}
		D^{\alpha }(uv)=\sum\limits_{\beta \leq \alpha }C_{\alpha }^{\beta
		}(D^{\beta }u)(D^{\alpha -\beta }v),
	\end{equation*}%
	where $\beta =(\beta _{1},\beta _{2},\beta _{3})$, $C_{\alpha }^{\beta
	}=C_{\alpha _{1}}^{\beta _{1}}C_{\alpha _{2}}^{\beta _{2}}C_{\alpha
		_{3}}^{\beta _{3}}$ is product of the binomial coefficients and $\beta \leq
	\alpha $ means that $\beta _{1}\leq \alpha _{1}$, $\beta _{2}\leq \alpha
	_{2} $, $\beta _{3}\leq \alpha _{3}$. Applying the differential operator $%
	D^{\alpha }$ with $|\alpha |=n$ to equation \eqref{14} and using the given
	above formula, we obtain 
	\begin{eqnarray}
		&&\varepsilon (\mathbf{x})\partial _{t}^{2}D^{\alpha }\mathbf{E}^{r}-\Delta
		D^{\alpha }\mathbf{E}^{r}+\sum\limits_{\beta \leq \alpha ,\,\beta \neq
			\alpha }C_{\alpha }^{\beta }\left( \partial _{t}^{2}D^{\beta }\mathbf{E}%
		^{r}\right) \left( D^{\alpha -\beta }\varepsilon (\mathbf{x})\right)  \notag
		\\
		&&+\sum\limits_{j=1}^{3}\sum\limits_{\beta \leq \alpha }C_{\alpha }^{\beta } 
		\left[ (\nabla D^{\beta }{E}_{j}^{r}))(\partial _{x_{j}}D^{\alpha -\beta
		}\varepsilon (\mathbf{x}))+(D^{\beta }{E}_{j}^{r})\nabla (\partial
		_{x_{j}}D^{\alpha -\beta }\ln \varepsilon (\mathbf{x}))\right]  \notag \\
		&=&D^{\alpha }\mathbf{F}^{r}(\mathbf{x},t).  \label{4.11}
	\end{eqnarray}%
	Calculating scalar product of both sides of equation \eqref{4.11} and $%
	2\partial _{t}D^{\alpha }\mathbf{E}^{r}$, we obtain the relation similar in
	the main part to \eqref{4.1}, namely: 
	\begin{eqnarray}
		&&\partial _{t}\left( \varepsilon |\partial _{t}D^{\alpha }\mathbf{E}%
		^{r}|^{2}+\sum\limits_{j=1}^{3}|\nabla D^{\alpha }{E}_{j}^{r}|^{2}\right) -2%
		\mathrm{div}\left( \sum\limits_{j=1}^{3}(\partial _{t}D^{\alpha }{E}%
		_{j}^{r})\nabla {E}_{j}^{r}\right)  \notag \\
		&&+2\sum\limits_{\beta \leq \alpha ,\,\beta \neq \alpha }C_{\alpha }^{\beta
		} \left[ (\partial _{t}D^{\alpha }\mathbf{E}^{r})\cdot \partial
		_{t}^{2}D^{\beta }\mathbf{E}^{r}\right] \left( D^{\alpha -\beta }\varepsilon
		(\mathbf{x})\right)  \notag \\
		&&-2\sum\limits_{i,j=1}^{3}\sum\limits_{\beta \leq \alpha }C_{\alpha
		}^{\beta }\left( \partial _{t}D^{\alpha }{E}_{i}^{r}\right) \left[ \left(
		\partial _{x_{i}}D^{\beta }{E}_{j}^{r}\right) \partial _{x_{j}}\ln
		\varepsilon (\mathbf{x})+\left( D^{\beta }{E}_{j}^{r}\right) \partial
		_{x_{i}x_{j}}D^{\alpha -\beta }\ln \varepsilon (\mathbf{x})\right]  \notag \\
		&=&2(\partial _{t}D^{\alpha }\mathbf{E}^{r})\cdot \mathbf{F}^{r}.
		\label{4.12}
	\end{eqnarray}%
	Integrating this identity over domain $\mathbb{R}_{t}^{4}$, $t\in (0,T]$, we
	arrive to the equality 
	\begin{eqnarray}
		&&\int\limits_{Y(t,T)}\left( \varepsilon (\mathbf{x})|\partial _{t}D^{\alpha
		}\mathbf{E}^{r}(\mathbf{x},t)|^{2}+\sum\limits_{j=1}^{3}|\nabla D^{\alpha }{E%
		}_{j}^{r}(\mathbf{x},t)|^{2}\right) d\mathbf{x}  \notag \\
		&=&-2\sum\limits_{\beta \leq \alpha ,\,\beta \neq \alpha }C_{\alpha }^{\beta
		}\int\limits_{\mathbb{R}_{t}^{4}}\left[ (\partial _{t}D^{\alpha }\mathbf{E}%
		^{r}(\mathbf{x},\tau ))\cdot \partial _{t}^{2}D^{\beta }\mathbf{E}^{r}(%
		\mathbf{x},\tau )\right] \left( D^{\alpha -\beta }\varepsilon (\mathbf{x}%
		)\right) \,d\mathbf{x}d\tau  \notag \\
		&&+2\sum\limits_{i,j=1}^{3}\sum\limits_{\beta \leq \alpha }C_{\alpha
		}^{\beta }\int\limits_{\mathbb{R}_{t}^{4}}\left( \partial _{t}D^{\alpha }{E}%
		_{i}^{r}(\mathbf{x},\tau )\right) \left( \partial _{x_{i}}D^{\beta }{E}%
		_{j}^{r}(\mathbf{x},\tau ))\right) \partial _{x_{j}}D^{\alpha -\beta }\ln
		\varepsilon (\mathbf{x})\,d\mathbf{x}d\tau  \notag \\
		&&+2\sum\limits_{i,j=1}^{3}\sum\limits_{\beta \leq \alpha }C_{\alpha
		}^{\beta }\int\limits_{\mathbb{R}_{t}^{4}}\left( \partial _{t}D^{\alpha }{E}%
		_{i}^{r}(\mathbf{x},\tau )\right) \left( D^{\beta }{E}_{j}^{r}(\mathbf{x}%
		,\tau ))\right) \partial _{x_{i}x_{j}}D^{\alpha -\beta }\ln \varepsilon (%
		\mathbf{x})\,d\mathbf{x}d\tau  \notag \\
		&&+2\int\limits_{\mathbb{R}_{t}^{4}}\partial _{\tau }D^{\alpha }\mathbf{E}%
		^{r}(\mathbf{x},\tau )\cdot D^{\alpha }\mathbf{F}^{r}(\mathbf{x},\tau )\,d%
		\mathbf{x}d\tau .  \label{4.13}
	\end{eqnarray}%
	Use now assumption \eqref{2.20}, \eqref{2.41} and the inequality $2\mathbf{%
		a\cdot b}\leq |\mathbf{a}|^{2}+|\mathbf{b}|^{2}$. Then, taking into account
	that $\sum_{\beta \leq \alpha }C_{\alpha }^{\beta }\leq 2^{3n}$ and $%
	C_{\alpha }^{\beta }\leq 2^{n}$, we obtain 
	\begin{eqnarray}
		&&\varepsilon _{0}\int\limits_{Y(t,T)}\left( |\partial _{t}D^{\alpha }%
		\mathbf{E}^{r}(\mathbf{x},t)|^{2}+\sum\limits_{j=1}^{3}|\nabla D^{\alpha }{E}%
		_{j}^{r}(\mathbf{x},t)|^{2}\right) d\mathbf{x}  \notag \\
		&\leq &C_{2}\int\limits_{\mathbb{R}_{t}^{4}}\left[ |\partial _{\tau
		}D^{\alpha }\mathbf{E}^{r}(\mathbf{x},\tau
		)|^{2}+\sum\limits_{j=1}^{3}|\nabla D^{\alpha }{E}_{j}^{r}(\mathbf{x},\tau
		)|^{2}\right] \,d\mathbf{x}d\tau  \notag \\
		&&+2^{n}\mu \sum\limits_{\beta \leq \alpha ,\,\beta \neq \alpha
		}\,\int\limits_{\mathbb{R}_{t}^{4}}\left( |\partial _{t}^{2}D^{\beta }%
		\mathbf{E}^{r}(\mathbf{x},\tau )|^{2}\,d\mathbf{x}d\tau
		+\sum\limits_{j=1}^{3}|\nabla D^{\beta }{E}_{j}^{r}(\mathbf{x},\tau
		)|^{2}\right)  \notag \\
		&&+\int\limits_{\mathbb{R}_{t}^{4}}|D^{\alpha }\mathbf{F}^{r}(\mathbf{x}%
		,\tau )\,d\mathbf{x}|^{2}d\tau ,  \label{4.15}
	\end{eqnarray}%
	where $C_{2}=2^{3n+1}\mu $.
	
	Since the relations $\beta \leq \alpha ,\,\beta \neq \alpha $ mean that $%
	|\beta |\leq |\alpha |-1=n-1$, then by the induction assumption \eqref{4.10}%
	, there exists a positive constant $C_{1}$ such that 
	\begin{equation*}
		2^{n}\mu \sum\limits_{\beta \leq \alpha ,\,\beta \neq \alpha }\,\int\limits_{%
			\mathbb{R}_{t}^{4}}\left( |\partial _{t}^{2}D^{\beta }\mathbf{E}^{r}(\mathbf{%
			x},\tau )|^{2}\,d\mathbf{x}d\tau +\sum\limits_{j=1}^{3}|\nabla D^{\beta }{E}%
		_{j}^{r}(\mathbf{x},\tau )|^{2}\right) \leq C_{1}M^{2}.
	\end{equation*}%
	Then we derive from \eqref{4.15} that 
	\begin{eqnarray}
		&&\int\limits_{Y(t,T)}\left( |\partial _{t}D^{\alpha }\mathbf{E}^{r}(\mathbf{%
			x},t)|^{2}+\sum\limits_{j=1}^{3}|\nabla D^{\alpha }{E}_{j}^{r}(\mathbf{x}%
		,t)|^{2}\right) d\mathbf{x}  \notag \\
		&\leq &C_{1}\int\limits_{\mathbb{R}_{t}^{4}}\left[ |\partial _{\tau
		}D^{\alpha }\mathbf{E}^{r}(\mathbf{x},\tau
		)|^{2}+\sum\limits_{j=1}^{3}|\nabla D^{\alpha }{E}_{j}^{r}(\mathbf{x},\tau
		)|^{2}\right] \,d\mathbf{x}d\tau  \notag \\
		&&+C_{1}M^{2}.  \label{4.16}
	\end{eqnarray}%
	Applying the Gronwall's inequality, we obtain 
	\begin{equation}
		\int\limits_{Y(t,T)}\left( |\partial _{t}D^{\alpha }\mathbf{E}^{r}(\mathbf{x}%
		,t)|^{2}+\sum\limits_{j=1}^{3}|\nabla D^{\alpha }{E}_{j}^{r}(\mathbf{x}%
		,t)|^{2}\right) d\mathbf{x}\leq C_{1}M^{2},~|\alpha |=n.  \label{4.17}
	\end{equation}
	
	Differentiating equation \eqref{4.11} $k\leq r-(n-1)$ times with respect to $%
	t$ and then calculating scalar product of both sides of the obtained
	equation and $2\partial _{t}^{k}\mathbf{E}^{r}$, we obtain relations %
	\eqref{4.11}-\eqref{4.17} with $\partial _{t}^{k}D^{\alpha }\mathbf{E}^{r}$
	instead $D^{\alpha }\mathbf{E}^{r}$. Therefore, the following estimates hold 
	\begin{equation*}
		\int\limits_{Y(t,T)}\left( \partial _{t}^{k+1}D^{\alpha }\mathbf{E}^{r}(%
		\mathbf{x},t)|^{2}+\sum\limits_{j=1}^{3}|\nabla \partial _{t}^{k}D^{\alpha }{%
			E}_{j}^{r}(\mathbf{x},t)|^{2}\right) d\mathbf{x}\leq C_{1}M^{2},~k\leq
		r-(n-1).
	\end{equation*}
	
	Thus, inequalities \eqref{4.10} hold with $n-1$ replaced by $n$. This
	justifies the mathematical induction method and we can set $n=r+2$ in %
	\eqref{4.10}. The latter proves the required inequalities \eqref{2.43}. $%
	\Box $

\end{document}

%% file: ex_shared.tex
\usepackage{amsmath, amssymb, latexsym, amscd,amsfonts,amstext}
\usepackage[mathscr]{eucal}
\usepackage{subfig}
\usepackage{graphicx}
\ifpdf
  \DeclareGraphicsExtensions{.eps,.pdf,.png,.jpg}
\else
  \DeclareGraphicsExtensions{.eps}
\fi

\newsiamremark{remark}{Remark}
\newsiamremark{hypothesis}{Hypothesis}
\crefname{hypothesis}{Hypothesis}{Hypotheses}

\headers{A SINGLE PDE FOR THE MAXWELL'S EQUATIONS}{V. G. Romanov and M. V. Klibanov}

\title{Can a Single PDE Govern Well the Propagation of the Electric Wave Field in a Heterogeneous Medium in 3D? \thanks{Submitted to the editors DATE.
\funding{The work of V.G. Romanov was supported by Mathematical Center in
	Akademgorodok at Novosibirsk State University (the agreement with Ministry
	of Science and High Education of the Russian Federation number
	075-15-2019-1613). The work of M.V. Klibanov was supported by US Army
	Research Laboratory and US Army Research Office grant W911NF-19-1-0044.}}}

\author{ Vladimir G. Romanov\thanks{Sobolev Institute of Mathematics, Novosibirsk 630090, Russian Federation, and Mathematical Center in Akademgorodok, Novosibirsk State University, Russian Federation. (\email{romanov{@}math.nsc.ru})}	
\and Michael V. Klibanov\thanks{Corresponding author. Department of Mathematics and Statistics, University of North Carolina at Charlotte, Charlotte, NC, 28223 USA  (\email{mklibanv@uncc.edu}).}
}
